\documentclass{emulateapj}

\def\gsim {\ifmmode {\buildrel>\over\sim}               % greater or similar
	\else {\lower.6ex\hbox{$\buildrel>\over\sim$}}\fi}
\def\lsim {\ifmmode {\buildrel<\over\sim}               % less or similar
	\else {\lower.6ex\hbox{$\buildrel<\over\sim$}}\fi}

\slugcomment{Not to appear in Nonlearned J., 45.}

\shorttitle{PDR chemistry in M82}
\shortauthors{Fuente et al.}

\begin{document}


\title{Photon-Dominated Chemistry in the Nucleus of M82:\\
Widespread HOC$^+$ emission in the inner 650 pc disk}


\author{A. Fuente$^1$,S. Garc\'{\i}a-Burillo$^1$,M. Gerin$^2$,D. Teyssier$^{3}$, 
A. Usero$^{1,4}$, J.R. Rizzo$^{1,5}$,P. de Vicente$^1$}
%\affil{Observatorio Astron\'omico Nacional (OAN), Apdo. 112,
%E-28800 Alcal\'a de Henares (Madrid,Spain)}
%
%\author{M. Gerin}
%\affil{Laboratoire d'Etude du Rayonnement et de la Matiere, UMR 8112, CNRS, Ecole
%Normale Superieure et Observatoire de Paris, 24 rue Lhomond, 75231 Paris Cedex 05, France}
%\email{aastex-help@aas.org}
%
%\author{J.R. Rizzo$^1$}
%\affil{Departamento de F\'{\i}sica, Universidad Europea de Madrid, Urb. El Bosque, E-28670 Villaviciosa 
%de Od\'on (Madrid, Spain)}
%\email{aastex-help@aas.org}
%
%\author{D. Teyssier$^2$}
%\affil{Space Research Organization Netherlands, PO Box 800, 9700 AV Groningen, The Netherlands}
%
%\author{A. Usero}
%\affil{Observatorio Astron\'omico Nacional (OAN), Apdo. 112,
%E-28800 Alcal\'a de Henares (Madrid,Spain)}
%
%\and
%
%\author{P. de Vicente}
%\affil{Observatorio Astron\'omico Nacional (OAN), Apdo. 112,
%E-28800 Alcal\'a de Henares (Madrid,Spain)}


\altaffiltext{1}{Observatorio Astron\'omico Nacional (OAN), Apdo. 112,
E-28803 Alcal\'a de Henares (Madrid,Spain)}
%\altaffiltext{2}{Laboratoire d'Etude du Rayonnement et de la Matiere en la Astrophysique, UMR 8112, CNRS, Ecole
%Normale Superieure et Observatoire de Paris, 24 rue Lhomond, 75231 Paris Cedex 05, France}
\altaffiltext{2}{LERMA, UMR 8112, CNRS, Ecole
Normale Superieure et Observatoire de Paris, 24 rue Lhomond, 75231 Paris Cedex 05, France}
%\altaffiltext{3}{Space Research Organization Netherlands, PO Box 800, 9700 AV Groningen, The Netherlands, under
%an external fellowship}
\altaffiltext{3}{SRON, PO Box 800, 9700 AV Groningen, The Netherlands, under
an external fellowship}
\altaffiltext{4}{Instituto de Astrof\'{\i}sica de Andaluc\'{\i}a (IAA),CSIC,Apdo. 3004, E-18080 Granada (Spain)}
%\altaffiltext{5}{Departamento de F\'{\i}sica, Universidad Europea de Madrid, Urb. El Bosque, E-28670 Villaviciosa 
%de Od\'on (Madrid, Spain)}
\altaffiltext{5}{Departamento de F\'{\i}sica, UEM, Urb. El Bosque, E-28670 Villaviciosa 
de Od\'on (Madrid, Spain)}


\begin{abstract}
The nucleus of M82 has been mapped in several 3mm and 1mm lines of 
CN, HCN, C$_2$H, c-C$_3$H$_2$, CH$_3$C$_2$H, HC$_3$N  
and HOC$^+$ using the IRAM 30m telescope. 
These species have been purposely selected as good tracers of photon-dominated 
chemistry. We have measured [CN]/[HCN]$\sim$5 in the inner 650~pc galaxy disk. 
Furthermore, {\bf we have detected the HOC$^+$ 1$\rightarrow$0 line with an intensity 
similar to that of the H$^{13}$CO$^+$ 1$\rightarrow$0 line. 
This implies a [HCO$^+$]/[HOC$^+$] ratio of $\sim$ 40.} {\bf These results corroborate the existence 
of a giant photo-dissociation region (PDR) in the nucleus of M82.}
In fact, the low [HCO$^+$]/[HOC$^+$] ratio can only be explained if the nucleus of M82 is formed by small
(r$<$ 0.02-0.2 pc) and dense (n$\sim$ a few 10$^4$--10$^5$ cm$^{-3}$) clouds immersed in an intense UV field
(G$_0$$\sim$ 10$^4$ in units of the Habing field). 
The detection of the hydrocarbons c-C$_3$H$_2$ and CH$_3$C$_2$H in the
nucleus of M82 suggests  that a complex carbon chemistry is developing in this 
giant PDR. 

\end{abstract}


%% subject header.  Objects should be in the appropriate "individual"
%% headers (e.g. quasars: individual, stars: individual, etc.) with the
%% additional provision that the total number of headers, including each
%% individual object, not exceed six.  The \objectname{} macro, and its
%% alias \object{}, is used to mark each object.  The macro takes the object
%% name as its primary argument.  This name will appear in the paper
%% and serve as the link's anchor in the electronic edition if the name
%% is recognized by the data centers.  The macro also takes an optional
%% argument in parentheses in cases where the data center identification
%% differs from what is to be printed in the paper.



\section{Introduction}
M82 is one of  the nearest and brightest starburst galaxies. Located at a distance of 3.9~Mpc, and with a luminosity 
of 3.7$\times$10$^{10}$~L$_{\odot}$, it has been extensively studied in many molecules. Compared to other prototypical 
nearby starburst galaxies like NGC~253 and IC~342, M82 presents systematically low abundances of the molecules 
NH$_3$, CH$_3$OH, CH$_3$CN, HNCO and SiO \citep{tak03}. 
Different explanations have been proposed to account for this peculiar chemistry. Since
all these molecules are related to dust grain chemistry, \citet{tak03} proposed that the 
formation of molecules on dust and/or evaporation to the gas phase is not efficient in M82.

On the other hand, several studies have revealed that the starburst has heavily influenced the
interstellar medium in M82 by producing high cosmic rays 
and UV fluxes. A low density ionized component is filling a substantial fraction
of the volume in M82 (see e.g Seaquist et al. 1996). 
The molecular gas
is embedded in this component in the form
of warm (T$_k$$>$50 K) and dense ($n$$>$10$^4$ cm$^{-3}$) clouds 
\citep{mao00}. Dense PDRs
are expected to form in the borders of these clouds \citep{wol90,lor96}.
\citet{wol90} have modeled the CII, SiII and OI emission and
estimated a UV field of G$_0$=10$^4$ in units of the Habing field
and a density of $n$$\sim$10$^5$ cm$^{-3}$ for
the atomic component.  \citet{sch93} reported
observations of the CI ($^3$P$_1$-$^3$P$_0$) line deriving a [CI]/CO column
density ratio ($\sim$ 0.5) higher than that observed in non-starburst galaxies
(e.g. [CI]/CO $\sim$ 0.15 in our Galaxy). They proposed that the
enhanced cosmic ray flux supplied by supernova remnants could  produce the
enhanced CI emission in M82. 
Recent results suggest that the strong UV flux dominates the
chemistry of the molecular gas in M82. \citet{gar02} obtained
a high-angular-resolution image showing widespread HCO emission in this galaxy. The
enhanced HCO abundance ([HCO]/[H$^{13}$CO$^+$]$\sim$3.6)
measured across the whole M82 disk was interpreted in terms of a giant
PDR of 650 pc size. 

In this paper, we present observations of a selected set of radicals (CN, C$_2$H), reactive
ions (CO$^+$, HOC$^+$) and small hydrocarbons (c-C$_3$H$_2$) which are 
excellent probes of the atomic to molecular transition in PDRs.
In particular, the [CN]/[HCN] ratio has been 
successfully used as a PDR indicator in regions with very different physical 
conditions in our Galaxy \citep{fue93,fue95,fue96}. 
%We use this ratio as a PDR tracer in the nucleus of M82.  
The detection of the reactive ion HOC$^+$ is 
almost unambiguously associated to regions with a high ionizing
flux, either PDRs or XDRs \citep{fue03,riz03,use04}. 
Recent works have revealed that the abundances of some
hydrocarbons, such as c-C$_3$H$_2$ and C$_4$H,
are an order of magnitude larger in PDRs than those predicted by gas-phase models
\citep{tey04,pet04}.  
%{\bf The failure of chemical models in explaining the hydrocarbons abundances in PDRs, as well as 
%the spatial coincidence between the distribution of small hydrocarbons and PAHs in these regions, 
%have led several authors\citep{fos03,fue03,tey04} to propose a formation process of small hydrocarbons linked to
%the PAHs chemistry.}  M82 is one of the starburst galaxies with most intense emission 
%in the PAHs infrared bands \citep{nor95,for03}, and therefore a good site for the study of
%these compounds. 

\section{Observations and analysis}
The observations were carried out  in June and November 2004 with the IRAM 30 m radiotelescope 
at Pico de Veleta (Spain). We used 2 SIS receivers tuned in single-sideband mode in the 1 mm and 3 mm bands. 
The observed transitions are: CN 1$\rightarrow$0 (113.490 GHz), CN 2$\rightarrow$1 (226.874 GHz),
HCN 1$\rightarrow$0 (88.631 GHz), C$_2$H 1$\rightarrow$0 (87.317 and 87.402 GHz), 
c-C$_3$H$_2$ 2$_{1,2}$$\rightarrow$1$_{0,1}$ (85.339 GHz), c-C$_3$H$_2$
6$_{1,6}$$\rightarrow$5$_{0,5}$ (217.822 GHz), CH$_3$C$_2$H 5$_k$$\rightarrow$4$_k$ (85.457 GHz), 
HC$_3$N 9$\rightarrow$8 (81.881 GHz) and HOC$^+$ 1$\rightarrow$0 (89.487 GHz). 
The line intensities have been
scaled to the main beam brightness temperature. The half-power beam width (HPBW) of the telescope 
is 29$''$ at 85 GHz, 22$''$ at 115 GHz and 12$''$ at 230 GHz.
We observed three positions across the M82 disk: the nucleus (RA(2000): 09$^h$55$^m$51.9$^s$,  
Dec(2000): 69$^{o}$4$'$47.11$''$) and the two peaks in the HCO emission [offsets ($+$14$''$,$+$5$''$) 
and ($-$14$''$,$-$5$''$) hereafter referred to as E. and W. knot respectively]. The observed spectra 
are shown in Fig. 1. To compare the 1.3mm and 3mm CN and c-C$_3$H$_2$ lines and make 
excitation calculations, we have derived beam filling factors from the  
H$^{13}$CO$^+$ interferometric image of \citet{gar02}. This assumption is justified 
since H$^{13}$CO$^+$ is an optically thin tracer of dense 
gas (n$>$ 10$^4$ cm$^{-3}$) and presents a quite uniform abundance in a wide range of physical conditions. 

The CH$_3$C$_2$H and C$_2$H column densities have been estimated using the 
LTE approximation.  For C$_2$H, we have assumed the typical rotation temperature 
in galactic PDRs, T$_{rot}$ = 10 K \citep{fue03}. 
In the case of CH$_3$C$_2$H, all the lines of the K-ladder are blended. 
In our estimates, we have assumed that 40\% of  the emission comes from the K=0 line and T$_{rot}$= 20 K.
These assumptions are based on the CH$_3$C$_2$H observations
of Sgr B2 and Orion which have similar  physical conditions to M82 \citep{chu83}. 
%The line parameters for this calculation have been taken from the JPL line catalog \citep{pic98}.
An LVG code has been used to derive densities and beam averaged column densities for the other
species. In these calculations, we have assumed T$_k$=60 K inferred 
from the NH$_3$ lines by \citet{wei01a}.

\section{Results}

We have estimated the hydrogen densities in the M82 disk
by fitting with a LVG code the  CN and C$_3$H$_2$ lines.
Hydrogen densities between  $n_{H_2}$$\sim$5$\times$10$^4$ cm$^{-3}$ 
and $n_{H_2}$$\sim$2$\times$10$^{5}$~cm$^{-3}$ 
are derived from both molecules. Assuming $n_{H_2}$$\sim$1$\times$10$^5$~cm$^{-3}$, we
have calculated the column densities averaged in a beam of 29$''$. 
The CN and HCN column densities
are quite constant along the galaxy disk with values, N(CN)=2.0$\pm$0.5$\times$10$^{14}$~cm$^{-2}$ and 
N(HCN)=4.0$\pm$0.5$\times$10$^{13}$ cm$^{-2}$. The [CN]/[HCN] ratio is similar
to 5 in all positions (see Table 1). As discussed in Section 4, this large value 
of the [CN]/[HCN] ratio is only reached in the most heavily UV exposed layers of a PDR.
The derived c-C$_3$H$_2$ and CH$_3$C$_2$H column densities
are also quite uniform  along the disk with  values,
N(C$_3$H$_2$)$\sim$1.7$\pm$0.4$\times$10$^{13}$ cm$^{-2}$
and N(CH$_3$C$_2$H)$\sim$1.0$\pm$0.6$\times$10$^{14}$ cm$^{-2}$. These column 
densities are in agreement with previous estimates by \citet{oik04}
and \citet{mau91}.
Finally, we have not detected the HC$_3$N 9$\rightarrow$8
line towards any position with an upper limit to the column density
N(HC$_3$N)$<$2$\times$10$^{12}$~cm$^{-2}$ towards the E. knot.
As discussed in Section 4 the non-detection of HC$_3$N and the
derived lower limit to the [c-C$_3$H$_2$]/[HC$_3$N] ratio
argue in favor of a PDR chemistry in M82.

We have detected the reactive ion HOC$^+$ in the
three selected positions across the M82 disk.  Furthermore,  the intensities of
the  HOC$^+$~1$\rightarrow$0 lines are similar, even larger, than those of 
the H$^{13}$CO$^+$~1$\rightarrow$0 lines.
Assuming the same physical conditions for H$^{13}$CO$^+$ and HOC$^+$,
we derive a [HCO$^+$]/[HOC$^+$] ratio of
$\sim$40 across the 650 pc inner disk, which 
is two orders of magnitude lower than that 
found in galactic giant molecular clouds (GMCs) \citep{app97}.
However, the H$^{13}$CO$^+$ spectra are derived from interferometric data. 
In order to asses the amount of missed
flux (and hence the H$^{13}$CO$^+$ column density), we have compared our results
to those of HCO$^+$ from \citet{ngu92}, which were obtained by single-dish observations. 
From our H$^{13}$CO$^+$ spectra, we obtain
N(H$^{13}$CO$^+$)$\sim$5$\times$10$^{11}$~cm$^{-2}$ across the galaxy.
Assuming  a $^{12}$C/$^{13}$C ratio of 89, this implies 
N(HCO$^+$)$\sim$4.5$\times$10$^{13}$~cm$^{-2}$. This value is in agreement within 
a factor of 1.5 with previous estimate by \citet{ngu92}.
Thus, even in the most conservative case we can conclude that  the [HCO$^+$]/[HOC$^+$] 
ratio is $<$ 80 in the M82 disk. Such low values of the [HCO$^+$]/[HOC$^+$] ratio have 
only been found in the galactic 
reflection nebula NGC~7023 \citep{fue03} and in the active nucleus of NGC~1068 
\citep{use04} and put severe constraints to the ionization degree of the molecular gas
in M82.

\section{Photon-dominated chemistry in M82}

\citet{ste95} found from detailed modeling that [CN]/[HCN]$>$1 arises naturally
in the surface layers (A$_v$$<$4 mag) of dense PDRs. 
Values of the [CN]/[HCN] ratio between 1--3  have been 
found in prototypical dense galactic PDRs 
\citep{fue93,fue95,fue96}. We have derived [CN]/[HCN] ratios $\sim$5 in all positions 
across the M82 nucleus suggesting that the molecular clouds
in this galaxy are bathed in an intense UV field. To determine the averaged
physical conditions of these clouds, we have carried out
model calculations using the plane-parallel PDR model developed 
by Le Bourlot and collaborators  \citep{leb93}. The model 
includes 135 species (HOC$^+$, CH$_3$C$_2$H and
HC$_3$N are not included) and standard gas-phase reactions.
Adopting G$_0$$\sim$10$^4$ \citep{wol90} and a total hydrogen nuclei 
density $n_H=n_{HI}+ 2 \times n_{H_2}$=4$\times$10$^5$~cm$^{-3}$, 
the model predicts that  [CN]/[HCN] ratios $\gsim$5 are 
only expected in regions at A$_v$$<$5--6 mag (see left panels of Fig. 2). 
This implies an important limit to the cloud sizes in the nucleus of M82.  
Since the clouds are bathed in a pervasive UV field, 
the averaged column density of individual clouds should be
$N_{H_2}$$\sim$10$^{22}$ cm$^{-2}$ in order to have averaged [CN]/[HCN] ratios
$\sim$5. 

Strong constraints are also derived from the HOC$^+$ observations. We 
have measured a [HCO$^+$]/[HOC$^+$] ratio of $\sim$ 40 in
the M82 disk. A simple calculation of the CO$^+$/HCO$^+$/HOC$^+$ 
chemical network shows that a high ionization degree, X(e$^-$)$>$ 10$^{-5}$, 
is required to have [HCO$^+$]/[HOC$^+$]$<$80 \citep{use04}. Our PDR model
shows that this high electron abundance is only found at 
A$_v$$<$4~mag (see left panels of Fig. 2). This implies that the averaged column
density of the clouds in the nucleus of M82 is N(H$_2$)$<$8$\times$10$^{21}$~cm$^{-2}$.  
Thus, the M82 nucleus seems to be formed by small (r$\sim$0.02-0.2 pc) and dense 
(n$\sim$10$^4$--10$^5$~cm$^{-3}$) clouds  immersed in a 
UV field of G$_0$$\sim$10$^4$. 
\citet{mao00} estimated similar column densities for the clouds in the
inner 400 pc disk of M82 by  modeling of the $^{12}$CO, $^{13}$CO and C$^{18}$O lines.  
Low values of the [HCO$^+$]/[HOC$^+$] ratio are also found in
XDRs with high ionization degree, such us active galactic nuclei (AGNs) \citep{use04}. 
However, the PAH emission is very low in AGNs since the doubly ionized
PAHs produced by X-rays are very unstable \citep{lea89}. The intense PAH emission 
observed in M82 \citep{nor95,for03} indicates that X-rays are not the driving agent of
the molecular gas chemistry in this galaxy.

\section{Hydrocarbon chemistry}

The hydrocarbon chemistry in PDRs has been a subject of increasing
interest both from the theoretical and the observational point of view.
Recent works have revealed that the c-C$_3$H$_2$ 
abundance in PDRs is similar to that in dark clouds. This is
quite surprising taking into account that this carbon cycle is easily photodissociated. In fact, PDR
gas-phase models fall short of explaining the observed
c-C$_3$H$_2$ and C$_4$H abundances by an order of magnitude \citep{tey04,pet04}. 
This is clearly seen when one
compares the [c-C$_3$H$_2$]/[HC$_3$N] ratio in PDRs and dark clouds (see Table 1).
While both species have similar abundances in dark clouds, the [c-C$_3$H$_2$]/[HC$_3$N] 
ratio is $>$10 in PDRs. Since both molecules are easily destroyed by photodissociation,
this suggests the existence of an additional c-C$_3$H$_2$ formation mechanism in PDRs. Several
authors have proposed formation processes of small hydrocarbons linked to the PAHs
chemistry \citep{tey04,pet04}. We have detected widespread emission of C$_2$H and c-C$_3$H$_2$
in the M82 disk. However, we have not  detected HC$_3$N towards any
position. The derived upper limits to the HC$_3$N column density in M82
show that the [c-C$_3$H$_2$]/[HC$_3$N] ratio is at least a  factor of 10 larger
in this starburst galaxy than in the prototypical galactic dark cloud TMC~1, and
is similar to that found in dense galactic PDRs (see Table 1). This result reinforces 
the scenario where the inner 650 pc region of M82 is a giant PDR. 

In addition to C$_2$H and c-C$_3$H$_2$, we have detected CH$_3$C$_2$H in all the positions 
across the M82 disk. This detection is rather puzzling since this molecule is easily photodissociated in
regions exposed to intense UV fields and therefore its abundance is expected to be very low
in these regions. Furthermore, this molecule is not detected in prototypical galactic PDRs,
like the Horsehead, which is known to be specially rich in carbon compounds \citep{tey04,pet04}. 
One possibility is that the CH$_3$C$_2$H emission arises in a  population of clouds
similar to the galactic GMCs.
The molecule CH$_3$C$_2$H is quite abundant in GMCs, presenting similar abundances to those of
chemically related species like CH$_3$OH and CH$_3$CN.
%(CH$_3$C$_2$H/CH$_3$CN and 
%CH$_3$C$_2$H/CH$_3$OH ratios between 1 and 10 in GMCs). 
The same behavior is observed in other prototypical starburst galaxies 
like NGC 253 where [CH$_3$C$_2$H]/[CH$_3$OH]$\sim$1
and [CH$_3$C$_2$H]/[CH$_3$CN$]\sim$8. 
However, [CH$_3$C$_2$H]/[CH$_3$OH]$>$8 and
[CH$_3$C$_2$H]/[CH$_3$CN]$>$25 
in M82 \citep{mau91,hut97}, i.e., 
a factor 3--10 larger than in NGC 253. 
These abundance ratios show that the hydrocarbon chemistry in M82 is
very different from that of GMCs.

{\it We think that the over-abundance of CH$_3$C$_2$H is also related 
to the enhanced UV flux in this galaxy.} In fact, CH$_3$C$_2$H and CH$_3$C$_4$H has been
detected by \citet{cer01} in the C-rich proto-planetary nebula
CRL 618. This nebula is specially rich in complex carbon compounds, in particular,
polyynes (C$_{2n+2}$H$_2$) and methylpolyynes (CH$_3$C$_{2n}$H).
They argued that the formation of these complex molecules has started
by the photolysis of acetylene (C$_2$H$_2$) and methane (CH$_4$), 
both processes occurring in the inner PDR of the 
planetary nebula. The photo-erosion of PAHs
could also contribute to enhance the abundance of these molecules by
releasing small carbon chains and  acetylene to the gas phase. 

\section{A global scenario for the chemistry in M82}

The observations presented in this paper show that the strong UV flux 
drives the chemistry of the molecular gas 
in the M82 disk. In this Letter, we explore whether the PDR chemistry
can solely account for all the molecular abundances 
measured in this galaxy so far. 
For this aim, we compare 
the molecular abundances in M82 to those of the prototypical
starburst NGC 253. 
%Both galaxies present important chemical differences \citep{tak95,tak03}.
Molecules such as NH$_3$, CH$_3$OH, CH$_3$CN, SO, SiO and HC$_3$N
are clearly under-abundant in M82 compared to NGC 253 \citep{wei01a,hut97,mau91,tak95,gar00,gar01}. 
All these molecules are easily photodissociated
and are not expected in a PDR. Other species such as CS, CN, HCN and HCO$^+$  
present fractional abundances $\sim$ a few 10$^{-9}$ in M82 
which are very similar to those measured in NGC 253 \citep{mau89}.
These molecules belong to the reduced group of species which have
significant abundances at
A$_v$$<$5 mag in our PDR model (see left panels of Fig. 2).  However, our
model predicts abundances a factor of 10 lower than those observed.
Cosmic rays and/or shocks could
contribute to enhance the abundances of these species.

To investigate the possible effect of the enhanced cosmic ray flux on
the chemistry, we have repeated the model calculations
with an enhanced cosmic rays flux of $\zeta$=4$\times$10$^{-15}$~s$^{-1}$. This is
the value estimated by \citet{suc93} to account for the physical conditions
of the molecular gas in M82 and is, very likely, an upper limit to the
actual cosmic ray flux. The enhanced cosmic rays flux does not change significantly
the value of the [CN]/[HCN] ratio and  the ionization degree at  A$_v$$<$5~mag 
(see right panels of Fig. 2). However, the fractional
abundances of HCN, CN and CS averaged over the region at A$_v$$<$5~mag are
a factor of 10 larger than those obtained with the standard value  $\zeta$=5$\times$10$^{-17}$~s$^{-1}$.
Moreover, these values
are in reasonable agreement with the observational results in M82. Thus, by including the enhanced 
cosmic ray flux we get a better fit of the model to the observational results. However, the HCO$^+$ abundance
still remains a factor of 10 lower than the observed one.

Some of the underabundant species in M82
like SiO are well known shock chemistry tracers. This suggests that shocks are not the dominant
mechanism at work in the galaxy disk. The SiO map reported by \citet{gar01} 
show that shocks are not occurring in the galaxy disk but in the
disk-halo interface because of the hot gas coming out of the plane in the 
nucleus of M82. 
This is consistent with the interpretation of M82 as an evolved starburst
compared to NGC 253. While in NGC 253 the molecular chemistry is determined
by the shocks associated with the early stages of star formation, in M82 it is determined
by the UV radiation produced by evolved stars and the enhanced cosmic 
rays flux \citep{gar02}.

\acknowledgments

We are grateful to the technical staff in Pico de Veleta for their supporting during
observations. This paper has been partially funded by the Spanish MCyT 
under projects DGES/AYA2000-927, ESP2001-4519-PE, ESP2002-01693, 
AYA2002-01241 and AYA2003-06473. We are also grateful to Jacques Le Bourlot
for his important help with model calculations.

\begin{figure}
%\epsscale{0.4}
\epsscale{0.5}
\plotone{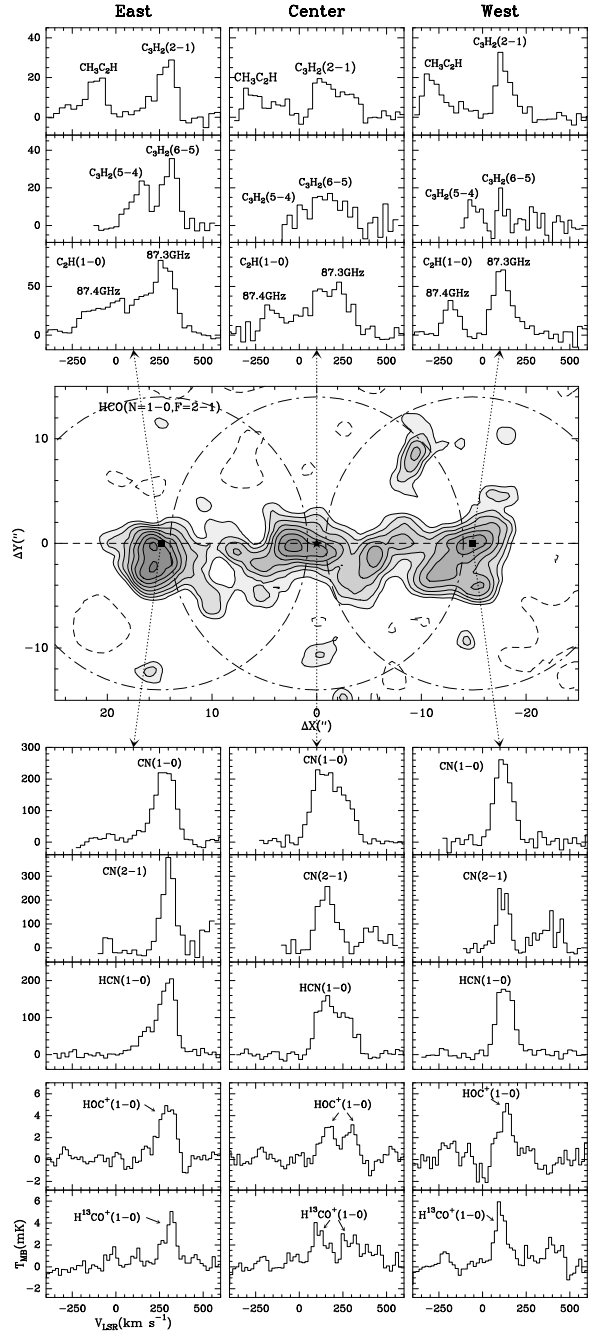}
\caption{Observed spectra towards the positions East  (E), Center and West (W) in M82. The 30m beam at 90 GHz
around the three observed positions has been drawn in the interferometrid HCO image by \citet{gar02}. 
We have fully sampled the HCO disk at 3mm.
\label{fig1}}
\end{figure}

\begin{figure}
%\epsscale{0.57}
\epsscale{0.7}
\plotone{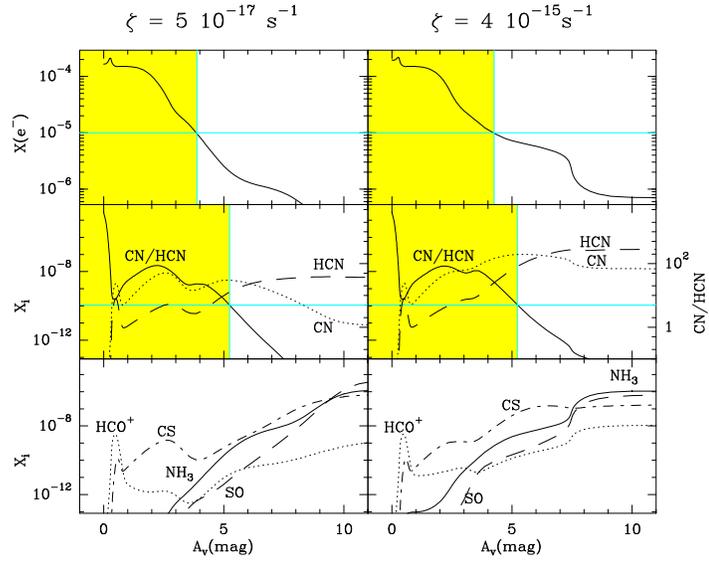}
\caption{Predictions for the abundances of various species derived from updated Le Bourlot et al.'s model. 
The calculations have been carried out for $n$=4$\times$10$^5$~cm$^{-3}$ and G$_0$=1$\times$10$^4$ 
in units of the Habing field. 
The cosmic ray flux is set to $\zeta$=5$\times$10$^{-17}$~s$^{-1}$ (galactic value) in left
panels and  $\zeta$=4$\times$10$^{-15}$~s$^{-1}$  (M82 value as derived by Suchkov et al. 1993)  in right panels. 
We have shadowed the
region of the plot in agreement with the observational results in M82. 
\label{fig2}}
\end{figure}

%\begin{figure}
%\includegraphics[angle=180,scale=.40]{histo.ps}
%\caption{Comparison between the molecular abundances in M82 (dark bar) and NGC 253.
%When possible, the fractional abundances have been derived from previously 
%published 30m data \citep{mau89,mau90,mau91,hut97} and those presented in this paper. 
%The assumed H$_2$ column densities averaged in a beam of $\sim$ 29$''$ are, 
%N(H$_2$) = 5.7 10$^{22}$ cm$^{-2}$ for M82
%and  N(H$_2$) = 1.1 10$^{23}$ cm$^{-2}$ for NGC 253 \citep{mau95}.
%Other references are:
%\citet{gar00} and \citet{gar01} for SiO abundances, \citet{tak95} for SO, \citet{mau03}and \citet{wei01a}
%for NH$_3$, and \citet{ngu92} for the HCO$^+$ and HCN abundances in NGC 253.}
%\end{figure}

\begin{table}
{\scriptsize
\begin{center}
\caption{Relative fractional abundances\label{tbl-3}}
\begin{tabular}{lcccccc|cc}
\tableline\tableline
Molecule                                   & \multicolumn{3}{c}{M82} & Orion Bar & NGC 7023 & Horsehead & TMC1 & L134N   \\ 
              & E.  & (0,0) & W.  &      IF          &   PDR peak                &  IR peak        &  CP         &      \\ \tableline
C$_2$H (10$^{13}$ cm$^ {-3}$)    & 58     & 41     & 30       & 57$^{(a)}$     & 3.8$^{(d)}$               & 16$^{(e)}$           & 7.2$^{(f)}$ & 15$^{(g)}$ \\ \tableline
c-C$_3$H$_2$/C$_2$H               & 0.02  & 0.04   & 0.02    & 0.02$^{(b)}$   & 0.03$^{(b)(d)}$        & 0.06$^{(e)}$        & 0.8$^{(g)}$ & 0.33$^{(g)}$  \\
CH$_3$C$_2$H/C$_2$H              & 0.2  & 0.4  & 0.3     & ...                 & ...                           & $<$0.07$^{(e)}$   & 1.4$^{(f)}$  & 0.07$^{(g)}$ \\
c-C$_3$H$_2$/HC$_3$N             & $>$ 6 & $>$ 4& $>3$   & $>$10$^{(b)(c)}$  & ...                     &  19$^{(e)}$         & 0.4$^{(h)}$  & 3$^{(g),(i)}$ \\ \tableline
CN/HCN                                     &   4     &   6    &   4      & 3$^{(a)}$        &   4--8$^{(d)}$         & ...                         & 0.07$^{(f)}$ & 0.07$^{(i)}$ \\ 
CN/HC$_3$N                              & $>$120  &  $>$58 & $>$80  & $>$ 180$^{(a)}$   & ...        &  ...                & 0.05$^{(f)(h)}$   & 0.9$^{(i)}$ \\ \tableline
HCO$^+$/HOC$^+$                    &  35         &  49      &  50   & $<$166$^{(a)}$    & 50--120$^{(b)}$  &  ...           & ... & ... \\
\tableline
\end{tabular}
%% Any table notes must follow the \end{tabular} command.
%\tablenotetext{a}{Assumed hydrogen density.}
\tablerefs{(a) Fuente et al.(1996); (b) Fuente et al.(2003); (c) Rodr\'{\i}guez-Franco et al.(1998); (d) Fuente et al.(1993);
(e) Teyssier et al.(2004); (f) Patrap et al.(1997); (g)Foss\'e (2003); (h) Takano et al. 1998; (i) Dickens et al. (2000).}
\end{center}
}
\end{table}

\end{document}